\documentclass[
	final,
	notitlepage,
	narroweqnarray,
	inline,
	twoside,
	]{ieee}

\usepackage{times}

\begin{document}

\title[A wireless physically secure \\key distribution system]{%
       A wireless physically secure \\key distribution system}

\author[BARBOSA]{Geraldo A. Barbosa$^*$
\authorinfo{$^*$G.\,A.\,Barbosa, QuantaSec -- Consulting and Projects in Physical Cryptography Ltd., Av. Portugal 1558, Belo Horizonte MG 31550-000 Brazil. E-mail: GeraldoABarbosa@gmail.com}}


\loginfo{Manuscript received June 8, 2016.}
\firstpage{1}
\maketitle
\newcommand{\be}{\begin{equation}}
\newcommand{\ee}{\end{equation}}
\newcommand{\bea}{\begin{eqnarray}}
\newcommand{\eea}{\end{eqnarray}}

\begin{abstract}
A fast and secure key distribution system is shown that operates in classical channels but with a dynamic protection given by the shot noise of light. The binary signals in the communication channel are protected by coding in random bases and by addition of physical noise that was recorded and added bit by bit to the signals. While the resulting signals are classical they carry the uncontrollable randomness information in the signal sent. The legitimate users start with a shared secret between them creating a measuring advantage over the adversary. This way the introduced noise does not affect the users but frustrates the attacker.
\end{abstract}


\section{Introduction}

A fast and secure {\em key distribution} system is presented to operate in generic communication channels, including wireless channels. The transmitted signals are deterministic (or perfectly copied) but include continuously recorded random noise that frustrates an attacker to obtain useful information. This noise affects the attacker but not the legitimate users that share an initial shared secret bit sequence $c_0$. The legitimate users will end up with a continuous supply of fresh keys that can be used even to encrypt information bit-to-bit in large volumes and fast rates.

The wireless  key distribution system discussed in this work uses the intrinsic light noise of a laser beam to frustrate an attacker to extract meaningful signals. However, this noise is not in the communication channel but it is recorded {\em before} reaching the channel.

Historically, cryptography using optical noise from coherent states in a communication channel can be traced back to \cite{grangier} and \cite{barbosa1}. The first uses quantum demolition measurements and quadrature measurements while the second uses direct measurements with no need for phase references or quantum features besides the presence of optical noise. The methods and techniques used are widely different. The use of the optical noise in this paper has a relationship to the one originally used in \cite{barbosa1}, where fiber optics communication in a noisy channel blocked information leakage to an adversary. An initial shared information on a $M$-ry coding protocol used by the legitimate users allowed them to extract more information from the channel than the one obtained by the adversary.  More recently that original idea was improved with a specific privacy amplification protocol \cite{Enigma2} while keeping the use of an optical communication channel.

The present work merges main ideas of the protection given by the light's noise in a protocol applied to wireless channels. Seed ideas on the use of a wireless channels using recorded physical noise were introduced from 2005 to 2007 \cite{deterministic noise}. This work brings those ideas of wireless channels secured by recorded optical noise to a practical level. It also opens up the possibility to immediate application of the technique to  mobile devices. This new scheme is detailed and the associated security level is calculated.
This system performs one-time-pad encryption with the securely distributed keys.

Symmetric keys with end-to-end encryption, where keys are kept secret by the users, may provide perfect secure communication for companies. Government distribution of keys for their users could guarantee secure communication among users as well as dispose of tools to access necessary exchanged information whenever a strong need exists. In the same way, companies that distribute keys for their users could comply   with legal requirements such as the All Writs Act (AWA) - as long as their key repository are kept under control.

A step-by-step description of this system will be made along this paper.
The key distribution system not just generates and distribute cryptographic keys but also provide functions like encryption and decryption between users (or ``stations") A and B: It is a {\em platform} for secure communications.

\section{Platform for secure communications}

\begin{figure}[h!]
\centerline{\scalebox{0.5}{\includegraphics{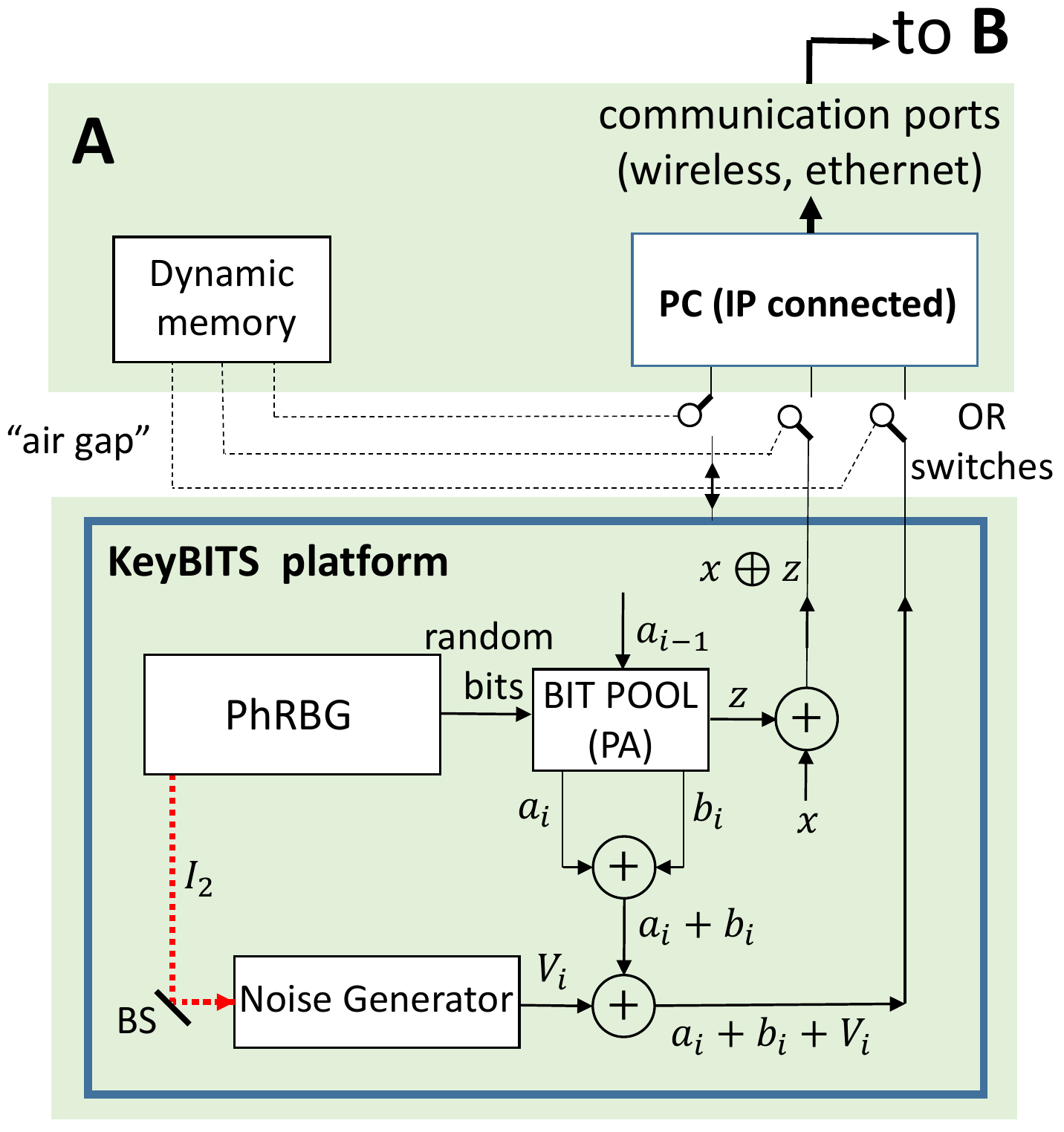}}}
\caption{ \label{wireless-PhRBGblocks}
Each station (A or B) is physically controlled by each user and is composed of an IP connected PC that exchange communications between stations. The platform has no direct connection to the communication channels.  Data flow in and out from the platform to the communication PC is done through a Dynamic memory and  OR switches. This memory contains instructions only allowing transit of authenticated packets with  fixed size. The platform contains the Physical Random Bit Generator (PhRBG), a Bit Pool and a Noise Generator to efficiently mimic the optical shot noise of a noisy optical communication channel.
}
\end{figure}

Fig.~\ref{wireless-PhRBGblocks} shows a block diagram of this platform for one of the users, say A. Users A and B have similar platforms.
Communication between A and B proceeds through the communication ports of a PC with access to the Internet (Top portion of Fig.~\ref{wireless-PhRBGblocks}). This PC works as the interface with the exterior and is isolated from the platform (Bottom of Fig.~\ref{wireless-PhRBGblocks}) by an air gap. In other words, the platform has no direct access to or from the Internet.  Data flow in and out  is done through a Dynamic memory conjugate with OR switches that only allow authenticated and fixed size packets.

The platform is roughly composed of two opto-electronic parts: 1) A fast Physical Random Bit Generator (PhRBG) and 2) a Noise Generator. The PhRBG delivers bits to a Bit Pool where a Privacy Amplification (PA) protocol is applied and encryption and decryption functions are performed. Description of these parts will be made along the paper; they are intertwined in their functionalities as as such their understanding are necessary for a full comprehension of the proposed system. A PC-mother board (not discussed in this paper) in the platform perform several operations and provide access for the users, including a graphical interface for platform control.

Although this is a quite general system allowing privacy in communications one could mention a few applications like secure communications for embassies or the secure transfer of large volumes of patient data among medical centers and insurance companies.

\section{Physical Random Bit Generator}
\label{sPhRBG}

The fast Physical Random Bit Generator (PhRBG) is of a novel type described in Ref.~\cite{Enigma1}. The PhRBG extract broad bandwidth fluctuations (shot-noise) of a laser light beam and delivers random voltage signals $(V_{+},V_{-})$ --signals that can be expressed as random bits-- to the Bit Pool.

\begin{figure}[h!]
\centerline{\scalebox{0.35}{\includegraphics{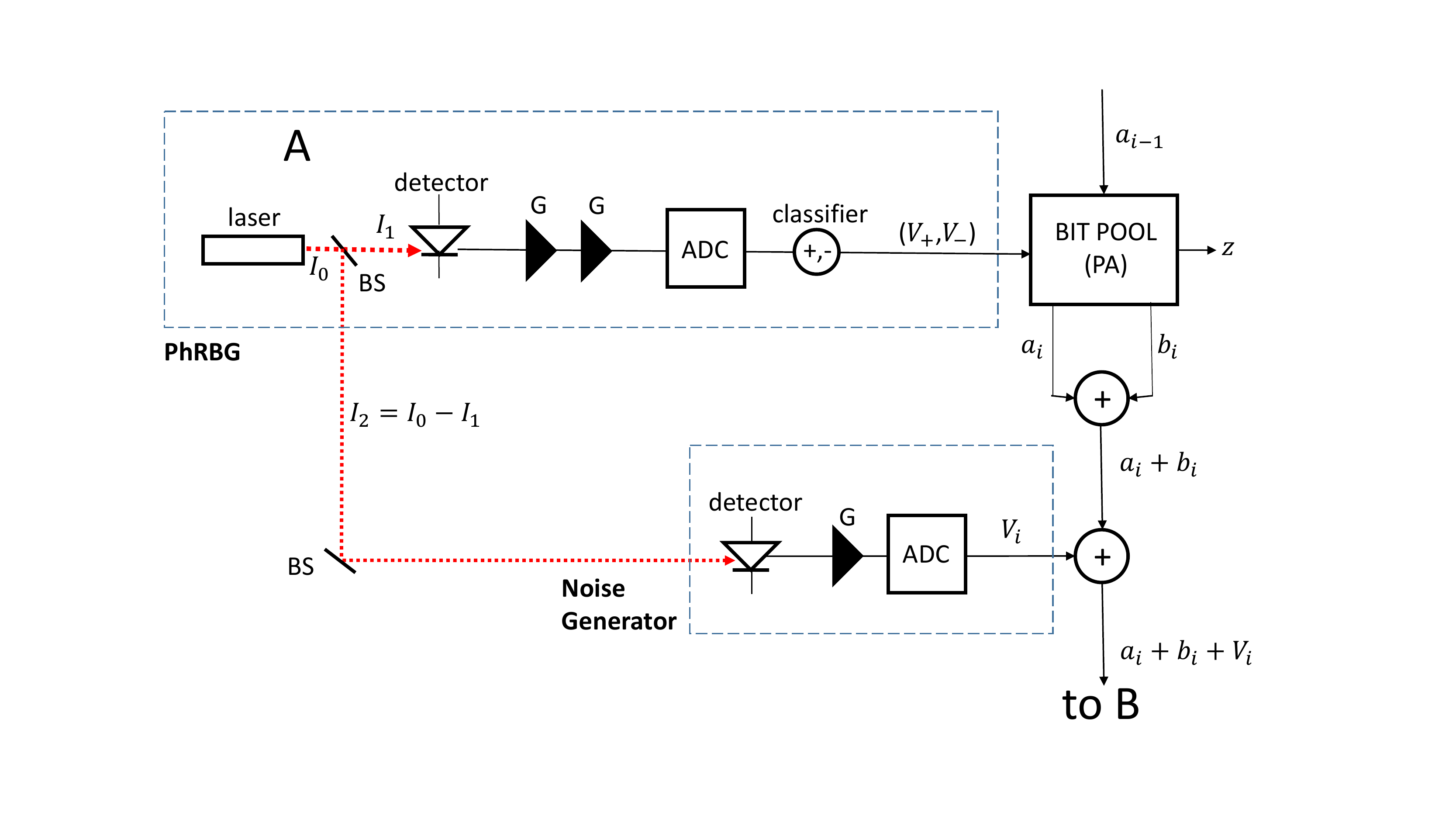}}}
\caption{ \label{wireless-PhRBG} Hardware to generate random bits (PhRBG), Noise Generator and bit pool for Privacy Amplification. The bit pool contain memories and a FPGA (Field Programmable Gate Array) to perform fast operations locally.}
\end{figure}

Fig.~\ref{wireless-PhRBG} provides more details. Left upper part of Fig.~\ref{wireless-PhRBG} shows the PhRBG.
 A laser beam excites a multi-photon detector and the voltage output pass through amplifiers G and an analog-to-digital (ADC) converter. The laser intensity $I_1$ and the gain G are adjusted to enhance the current from the noisy optical signals well above electronic noises:
 \bea
 \overline{(\Delta I_{light})^2} \gg \overline{(\Delta I_{electronic})^2}\:.
 \eea
 It also necessary to work below the range where the ratio noise/signal is too small. In terms of the number of photons $n$:
\bea
  \frac{ \mbox{Noise}}{ \mbox{Signal}}  = \frac{\sqrt{(\langle \Delta n \rangle })^2}{\langle n \rangle}=\frac{1}{\sqrt{\langle n \rangle}} \rightarrow \mbox{not small}\:.
 \eea

 In other words, the desired signals are optimized optical shot-noise signals that allow a good number of detection levels from an ADC. The stream of digitalized fluctuating signals are classified within short time intervals in signals above the average value as bit 1 signals ($V_{+}$) while signals below the average are identified as bit 0 signals ($V_{-}$).

 Sampling time for acquisition of the bit signals are set much shorter than the coherence time of the laser used. By doing so samplings occur within a fixed optical {\em phase} of the sampled photons. This leads to photon {\em number} fluctuations that are maximal: Although phase and number (or photon amplitude) are not strictly conjugate variables, there is an uncertainty relationship for number and phase.

The individual bit signals generated by the PhRBG around time instants $t_i$ will be designated by $a_i$ and a sequence of $a_i$ by $a$. Notation $a$ sometimes designates a sequence of bits or the size of this sequence whenever this does no give rise to notational problems.

User A wants to transmit in a secure way these random $a$ bits to user B.

Appendix \ref{KeyBITS/PhRBG} comments on the PhRBG.

\section{Bit Pool and Noise Generator}
\label{bitpool}

Fig.~\ref{wireless-PhRBG} also shows at the right upper side a Bit Pool where random bits generated from the PhRBG are stored together with bits $b_i$ that were already acquired and recorded. The initial sequence $\{b_0\}$ is taken from  a secret sequence $c_0$ of size $c_0=m a$ initially shared between the legitimate users. The Bit Pool outputs signals $b_i+ a_i$.
Bits $b_i$ act as a modulation or encryption signals to the random bits $a_i$.  After application of the PA protocol a final distillation of $z$ bits, over which the attacker has no knowledge, will be available for encryption and decryption purposes.
  A full discussion of these operations are made ahead when discussing the physical modulation of the signals and the Privacy Amplification protocol.

\subsection{Noise Generator and recorded optical shot-noise}

Bottom part of Fig.~\ref{wireless-PhRBG} shows the Noise Generator. A laser beam with intensity $I_2$ is detected, amplified to produce optical shot noise limited signals. These signals are digitalized producing a sequence of independent noise signals $V_N=\{V_i\}$. $V_i$ is added to the signals $a_i+b_i$ giving $a_i+b_i+V_i$ and sent to B. The noise contribution $V_N$ replaces the intrinsic optic noise in an optical channel. The magnitude and format of this noise will be shown after presenting the idea of $M-$ry bases.


The first modulation signal $b_0$ is defined by $m$ random bits from $c_0$. In general the modulating random signal $b_i$ can be seen as a transmission {\em basis} for $a_i$. One may as well see bits $a_i$ as a message and $b_i$ as an encrypting signal. To generate each $b_i$, or one number among $M$, $m$ bits are necessary ($m=\log_2 M$).

 It is to be understood that the $M-$ry coding interleaves bits in the sense that the same bit signal superposed to a basis $b_k$ representing a bit 1 (or 0) represents the opposite bit 0 (or 1) in a neighbor basis $b_{k-1}$ or $b_{k+1}$. For example, see Fig. 1 in \cite{Enigma2} for a physical representation of these interleaved bits in the optical phase space. Other possible realization of distinct neighboring levels with distinct bits could be made with levels separated by small physical displacements different from phase, e.g. amplitude, as shown in Fig.~\ref{wirelessLevels2}.  This, together with the added noise $V_N$ do not allow the attacker to obtain the bit $a_i$.

 \subsection{Noise Generator and $M-$ry bases}
 \label{VN}

As shown in the bottom right part of Fig.~\ref{wireless-PhRBG} a random signal $V_i$ is added to $a_i+b_i$, giving  $a_i+b_i+V_i$ to be sent to B. It is emphasized that although the signal sent from A to B is deterministic, $V_i$ is a {\em recorded random noise} that varies from bit to bit. A recorded signal is deterministic by definition because it can be perfectly copied. However, this recorded noise is an instance of an unpredictable event by nature.

In nature the noise intensity is continuous but the recorded digitalized noise is distributed among the $M$ levels supplied by an  Analog to Digital Converter. This statistical distribution among $M$ levels also has a characteristic deviation $\sigma_V$. This will be discussed ahead.

 \begin{figure}[h!]
\centerline{\scalebox{0.5}{\includegraphics{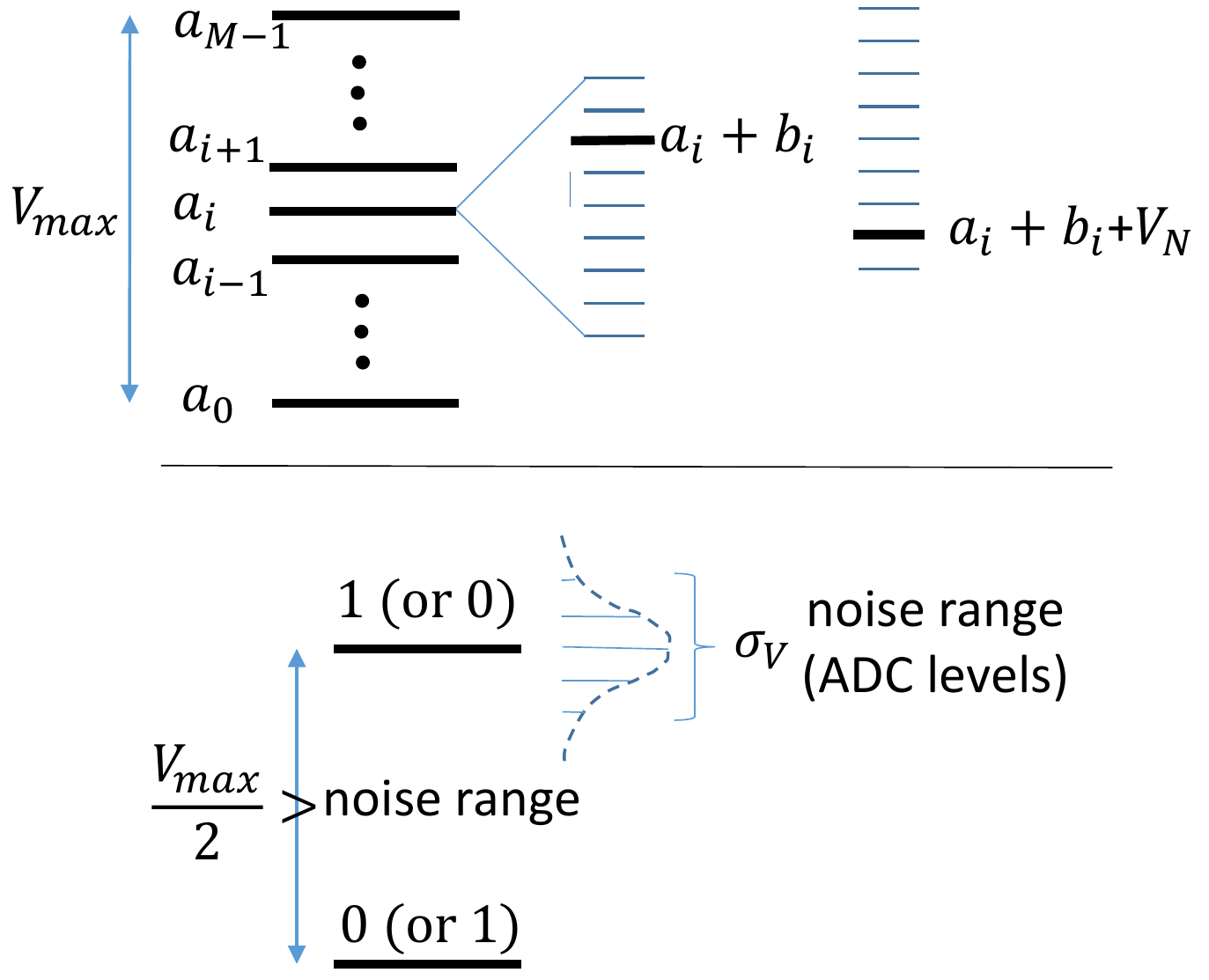}}}
\caption{ \label{wirelessLevels2} TOP - The physical amplitude signal representing a given basis $b_i$ is added to the signal representing a bit $a_i$. The basis signal is known to users A and B but not to the attacker. A signal $a_i$ is to be seen as a given bit in basis $b_i$ (1 or 0) but this same signal $a_i$ will be seen as the opposite bit (0 or 1) when attached to neighboring bases $b_{i+1}$ or  $b_{i-1}$. Therefore even a small noise $V_i$ added to $a_i+b_i$ do not allow an attacker to know which bases have been used. However, both A and B know $b_i$ and thus the bit sent $a_i$ can be extracted. Physically, distinct modulation voltage signals may represent bits and bases. There are $M$ bases; they can be assumed separated by $V_{max}/M$. A bit 1 and a bit 0 can be assumed physically separated by $V_{max}/2$. Voltage signals can be assumed cyclic in the sense that $V_{max}+\epsilon \rightarrow \epsilon$.
BOTTOM - The amplitude between a signal $a_i$ representing a bit and the opposite one is greater than the noise range. As voltage modulation of signals representing 1 and 0 are separated by $V_{max}/2$ this allows a precise bit determination by the legitimate users. On the contrary, the attacker struggles unsuccessfully with resolution of neighbouring levels.}
\end{figure}

   The noise signal $V_i$ is derived from a split beam of intensity $I_2$ (see left bottom part of Fig.~\ref{wireless-PhRBG}). Light from this derived beam excites a multi-photon detector, the output is amplified by  G and digitalized. An extra amplifier G may be adjusted to levels compatible to the signal $a_i+b_i$. In other words, the added noise  mix potential bases and bits so that the attacker could not identify either the bit or the basis sent. Fig.~\ref{wirelessLevels2} sketches the addition of the random basis $b_i$ and the added noise $V_i$. The attacker does not know either the basis $b_i$ neither the noise $V_i$ and, therefore, cannot deduce the bit sent $a_i$ from the total signal $a_i+b_i+V_i$.

\section{Privacy Amplification and Fresh Bit Generation by A and B}

The Privacy Amplification process to be utilized was first shown in Section IX of Ref.~\cite{Enigma2}; it utilized the formalism originally developed in Ref.~\cite{BBCM}. In the present work it is applied to a classical communication channel instead of a noisy fiber optic channel.

Briefly, the following steps are performed:

1) The Bit Pool starts with the bit sequence of size $c_0=ms$ (bits $b_i$) already shared by A and B. A sequence $a$ of  bits, $a=\{a_i\}$, is generated by the PhRBG and stored in the Bit Pool by user A.
The sequence $a$ is sent from A to B after the preparation that add bases and noise. A number of bits $a+m a$ is used for the task of creating bits $a_i$ and bases $b_i$ to be sent from A to B as  $\{a_i+b_i\}$.

2) An instance of a universal hash function $f$ is sent from A to B.

3) The probability for information leakage of bits obtained by the attacker over the sequence sent is calculated (as indicated ahead), generating the parameter $t$ (number of possibly leaked bits). In other words, from sequence $\{ 0,1  \}^n$ the attacker may capture $\{ 0,1  \}^t$.

The PA protocol includes the following steps:
From the $n=a+b$ bits stored in the Bit Pool,  $t$ bits are destroyed:
\bea
\{ 0,1  \}^n \rightarrow \{ 0,1  \}^{n-t}\:.
\eea
An extra number of bits $\lambda$ is reduced as a security parameter \cite{BBCM}. This reduction in bit numbers is then
\bea
\{ 0,1  \}^n \rightarrow \{ 0,1  \}^{n-t}\rightarrow \{ 0,1  \}^{n-t-\lambda}
\eea
where $\{ 0,1  \}^{n-t-\lambda}$ is the final number of bits.
The initial total amount of bits $n$ in the Bit Pool was then reduced to $r=n-t-\lambda$. These remaining bits are then further randomized by the PA protocol \cite{Enigma2}. The protocol establishes that the attacker has no information on these reduced and ``shuffled" number of bits $r$.

The number $r$ of bits can be rearranged in sizes as follows
 \bea
 r&=&n-t-\lambda=(a+b)-t-\lambda
 =(a-t-\lambda)+b \nonumber \\
 &=&(a-t-\lambda)+m a\equiv z+ ma \:.
 \eea
 The sequence of size $z\equiv (a-t-\lambda)$ (see output from Bit Pool in Fig.~\ref{wireless-PhRBG}) will be used as {\em fresh bits} for encryption while the sequence of size $ma$ will form the new bases $\{ b_i \}$ for the next round of bit distribution. The process can proceed without the legitimate users having to meet or use a courier to refresh an initial sequence $m a$. Other rounds then may proceed.

The PA theory \cite{BBCM} says that after  reducing the initial number of bits from $n=a+m a$ ($m a$ initially shared and $a$ fresh bits) to $r=n-t-\lambda$,  the amount of information that may be acquired by the attacker is given by the {\em Mutual Information} $I_{\lambda}$.
Corollary 5 (pg. 1920) in Ref.~\cite{BBCM},  gives the information leaked to the attacker:
\bea
\label{Ir}
I_{\lambda}=\frac{1}{\ln 2 \times 2^{\lambda}}=\frac{1}{\ln 2 \times 2^{n-t-r}}\:.
\eea

\subsection{Protocol steps}

Table \ref{PAtable} list all steps of the protocol. The basis assigned for each bit sent uses $\log_2 M$ to encode it and the process is continuously sustained in rounds of $s$ bits, in an unlimited way. This procedure has been shown to be very fast in hardware.
\begin{center}
{
\begin{table}
\caption{Privacy Amplification protocol for the \hskip3cm  wireless platform}
\label{PAtable}
\begin{tabular}{|l|c|c|}
  \hline
  \multicolumn{3}{|c|}{\small \bf PA protocol}\\
  \multicolumn{3}{|l|}{\small{\bf INITIALIZATION:} A and B share $c_0$ of size and entropy $m\: s$.}\\
  \hline
  \hline
  \multicolumn{3}{|l|}{\small\bf Station A}\\
  \hline
   {\small $\#$} &{\small ACTION}&{\small OBJECTIVE } \\
   \hline
  {\small 1a} &{\small  $a_i=\mbox{GetString(PhRBG)}$ }&{\small Get bitstring from PhRBG}\\
  {\small 1b} &{\small  $b_i=c_{i-1}[1,m s ]$ }&{\small Extract $m \: s$ from pool for bases $b$}\\
  {\small 1c }& {\small Code\&Send$(a_i,b_i)$ }&{\small Send over classical channel}\\
  \hline
  {\small 2} &{\small  Send $f$  }         &{\small Send instance of universal hash $f$}\\
  &&{\small over classical channel}\\
  \hline
  {\small 3a} &{\small $ c_i=f(c_{i-1}||a_i)$ }&{\small A applies PA from $ms+s$ bits }\\
  &  &{\small reducing them to $m s+s-t-\lambda$}\\
  {\small 3b}&{\small  $z_i=$} &{\small A uses $s-t-\lambda$}\\
                 &    {\small $\!\!\!\!c_i[ms+\!\!1,ms\!+s\!-\!t\!-\!\lambda]\!\!\!\!$}           &{\small bits from pool as the key stream $z$.}\\
                  &           &{\small The remaining $m\: s$ bits form }\\
                   &          &{\small the bases for next round.}\\
  \hline
  \hline
  \multicolumn{3}{|l|}{{\small \bf Station B}}\\
  \hline
 {\small 1a} &  &{\small no matching step to A's}\\
{\small  1b} & {\small $ b_i=c_{i-1}[1,m s]$}&{\small Get bases bits from initial pool value} \\
 {\small 1c} &{\small  $a_i=$Receive\&Decode$(b_i)$}&{\small Receive bits from classical channel} \\
 \hline
 {\small  2} & {\small  Receive $f$ }&{\small receive instance of universal hash $f$}\\
  \hline
{\small  3a} & {\small $c_i=f(c_{i-1}||a_i)$ }&{\small B applies PA from $ms+s$ bits} \\
  &  &{\small reducing them to $ms+s-t-\lambda$}\\
{\small  3b} &   {\small  $z_i=$}        &          {\small B uses $s-t-\lambda$}\\
      &        {\small $\!\!\!\!c_i[ms+\!\!1,ms\!+s\!-\!t\!-\!\lambda]\!\!\!\!$}               &{\small bits from pool as the key stream $z$.}\\
                  &           &{\small The remaining $m\: s$ bits form }\\
                   &          &{\small the bases for next round.}\\
  \hline
\end{tabular}
\end{table}
}
\end{center}

 A and B use the protocols in a concerted manner and extract a sequence $z$ of bits over which the attacker has no knowledge.  One should recall that the communication channel is classical and the signals contain recorded optical noise modulating each bit sent. At every round A and B know the basis used and they use this to their  advantage so that the noise $V_N$ does not disturb identification of $a_i$. A secure distilled stream of bits from A is transferred to B.

The protocols proceeds to other similar runs. After $n$ runs, Alice and Bob share $n z$ bits.

\section{Leakage probability and mutual information $I_{\lambda}$}

Calculation of the mutual information $I_{\lambda}$ that is directly connected to the probability for an attacker to extract useful information sent from A to B. It depends on the parameter $t$ (number of possibly leaked bits in a sequence sent).

In the wireless scheme the number of levels used as bases depends on the digital hardware utilized (8 bits resolution$\rightarrow M=256$, 10 bits resolution $\rightarrow M=1024$ and so on). This converter sets the maximum number of levels $M$. Voltage signals $V_k\:,(k=0,1, 2 \dots M)$ will represent these bases and to alternate bits in nearby bases one may chose bases by voltage values given by
\bea
V_k=V_{\tiny \mbox{max}}\left[ \frac{k}{M}+\frac{1-(-1)^k}{2} \right]\:.
\eea

At the same time, as voltage signals $V_N$ representing recorded optical noise will be added to these values, these values should have a span smaller than $V_{\tiny \mbox{max}}$ (see Fig.~\ref{wirelessLevels2}). However, this span must be large enough to cover a good number of bases so that the attacker cannot resolve the basis $b_i$ when a bit $a_i$ is sent.  The actual optical noise has a continuous span but the recorded region is set by digitalized levels of the ADC used. Setting the spacing of signals for bases similar to the spacing $V_{\tiny \mbox{max}}/M$ of recorded noise levels, one could set the digitalized noise deviation, by adjusting the gain  $G$, such that
\bea
V_{\tiny \mbox{max}}/M\ll \sigma_V\ll V_{\tiny \mbox{max}}\:.
\eea

This condition can be mapped to the same formalism utilized in the POVM (Positive Operator Valued Measure) calculation developed in \cite{barbosa1} and from which the leakage bit probability $t$ can be obtained. One may write the probability for indistinguishability between two levels separated by $\Delta k$, as
\bea
P_{\Delta k}=e^{-\frac{|\alpha|^2}{4} \left(\frac{V_{\Delta k}}{V_{\tiny \mbox{max}}}\right)^2}
=
e^{    -\frac{|\alpha|^2}{4} \frac{(\Delta k)^2}{M^2}   }
\equiv e^{- \frac{\Delta k^2}{2(\sigma_k)^2}     }\:.
\eea
The expected deviation $\sigma_k$ in the number of levels is
\bea
\sigma_k=\sqrt{\frac{2}{ \langle n \rangle}}M\:,
\eea
where $\langle n \rangle=|\alpha|^2$ and $\alpha$ is the coherent amplitude of a laser.

\begin{figure}[h!]
\centerline{\scalebox{0.3}{\includegraphics{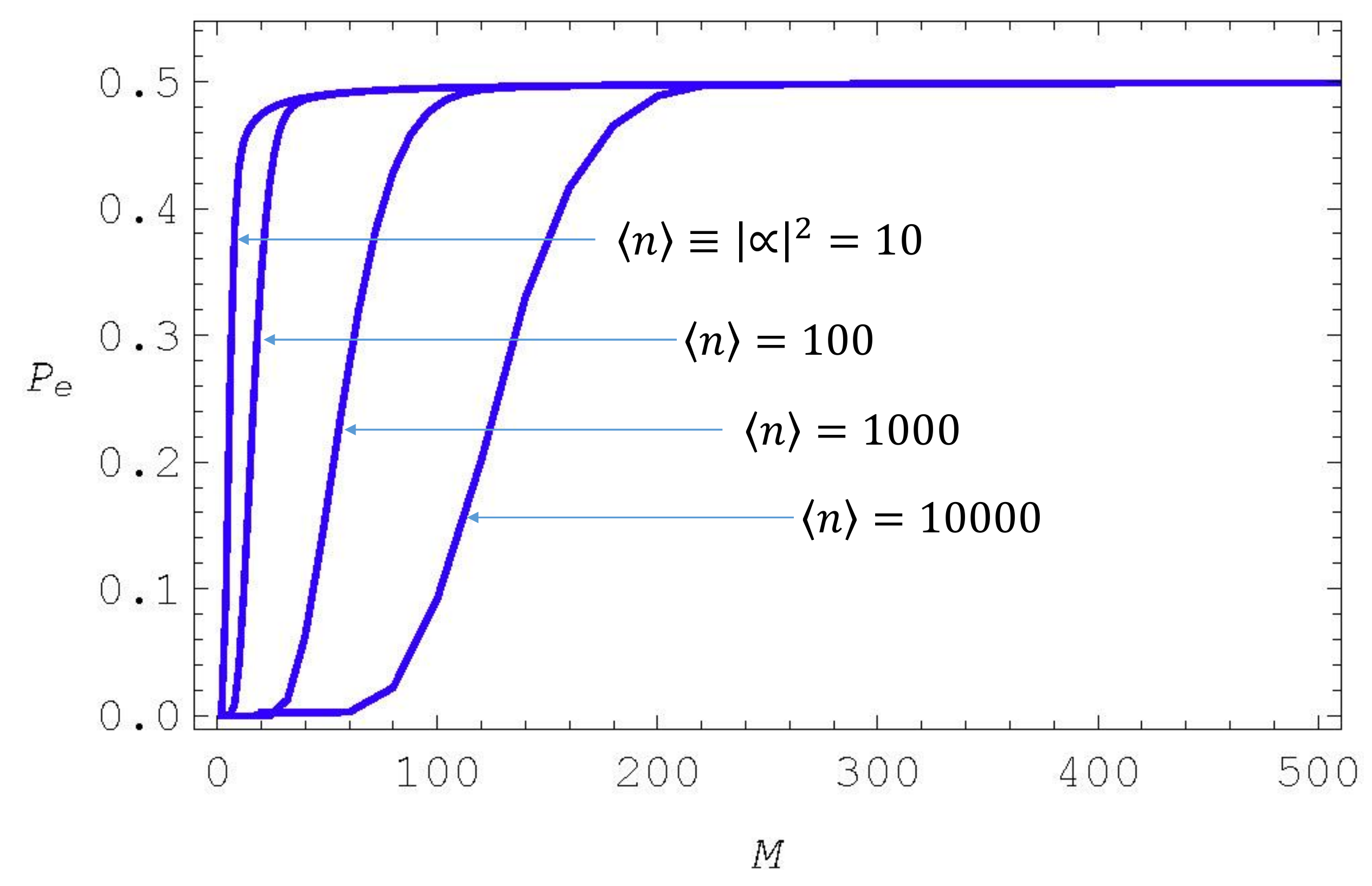}}}
\caption{ \label{pAll500alpha10to10000}Probability of error for an attacker on a bit  as a function of the number $M$ of bases used and the average number of photons $\langle n \rangle$ carrying a bit. }
\end{figure}

Calculation of the probability of error $P_e$ for an attacker to obtain a bit sent follows what was done in \cite{barbosa1}.
Fig.~\ref{pAll500alpha10to10000} exemplifies these errors for a set of $M$ values (number of bases) and number $\langle n \rangle$ of photons detected.
For a sequence of $s$ bits sent the parameter $t$ (bit information leaked in $s$) in Eq.~\ref{Ir} will be $t=(0.5-P_e)\times s$. With $t$ calculated and the safety parameter $\lambda$ defined, the probability for information that could be leaked to the attacker  is calculated.
It can be shown \cite{barbosa1} that $t \sim 10^{-4}$ can be easily obtained; therefore with a sequence os $s=10^6$ bits sent, this gives $t\sim 10^2$.

\begin{figure}[h!]
\centerline{\scalebox{0.45}{\includegraphics{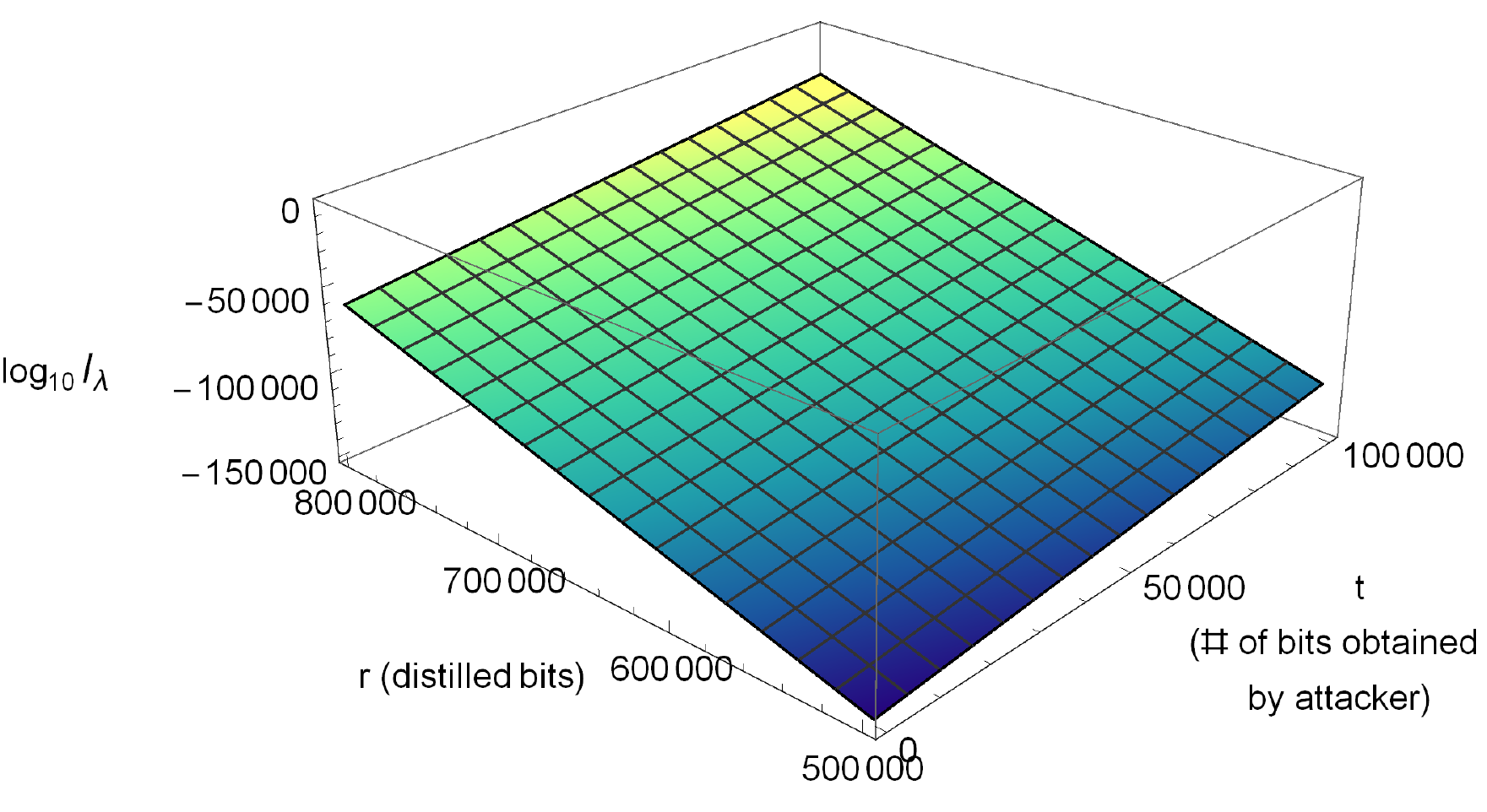}}}
\caption{ \label{plogI} $\log_{10}$ of the Mutual Information $I_{\lambda}$ leaked to the attacker after Privacy amplification is applied. In this example $10^6$ bits are sent. $t$ gives the number of bits leaked to the attacker before PA is applied and $r$ is the distilled or useful remaining bits.}
\end{figure}

Fig.~\ref{plogI} exemplifies the PA effect by ($\log_{10} I_{\lambda}$) (see Eq.~\ref{Ir}) as a function of $r$, $(0,1)^n\rightarrow(0,1)^r$, and  $t$, number of bits leaked to the attacker.

\section{Other application examples}
Above sections described the basic parts of the platform for secure communications.

The use of a FPGA (Field Programmable Gated Array) and memory allow functions like bit  storage and encryption and  perform  the ``Bit Pool" functions necessary. Custom tailored applications can also be programmed under this fast hardware processing.

The current rack holding the PhRBG  can be  reduced to a small size  with output directly coupled to a smart phone. This can provide true secure communications (bit-to-bit encryption) between cellular telephone users as another application example. It is also useful to call the attention to the reader that a decentralized protocol for bit-by-bit encryption for $N-$users exists \cite{jeroen}. Under this protocol, once $N$ users acquire a long stream of random bits from the PhRBG, they can exchange secure information among them without any need to contact a central station to synchronize their bit streams.

Both Secure Data and Voice Over Internet (VOIP) can be implemented. It should be emphasized that it is important that the key storage must be kept ``outside" of the mobile device and that the flow of information from the key generation and encrypting unit to the device connected to the Internet should be strictly controlled.

Other steps, more costly, can produce an ASIC (Application Specific Integrated Circuit) to reduce the system to a chip size device.

Another possible application example for the platform is to feed a Software-Defined-Radio (SDR) with cryptographic keys for bit-by-bit encryption/decryption capabilities. This could bring absolute security for Data and Voice communications through SDR.

\section{Conclusions}

It was shown how to achieve wireless secure communication at fast speeds with  bit-to-bit symmetric encryption. The hardware requirements was described and it was shown how to calculate the security level associated to the communication. Miniaturization steps may allow easy coupling to mobile devices.  The key storage have to be under control of the legitimate users and no key should ever be stored where a hacker could have command/control of the system. A correct implemented system would offer privacy at top-secret level for the users. Furthermore, the correct choice of parameters creates a post-quantum security privacy.

\appendix

\section{Platform - rack implementation}
\label{KeyBITS/PhRBG}

\begin{figure}[h!]
\centerline{\scalebox{0.2}{\includegraphics{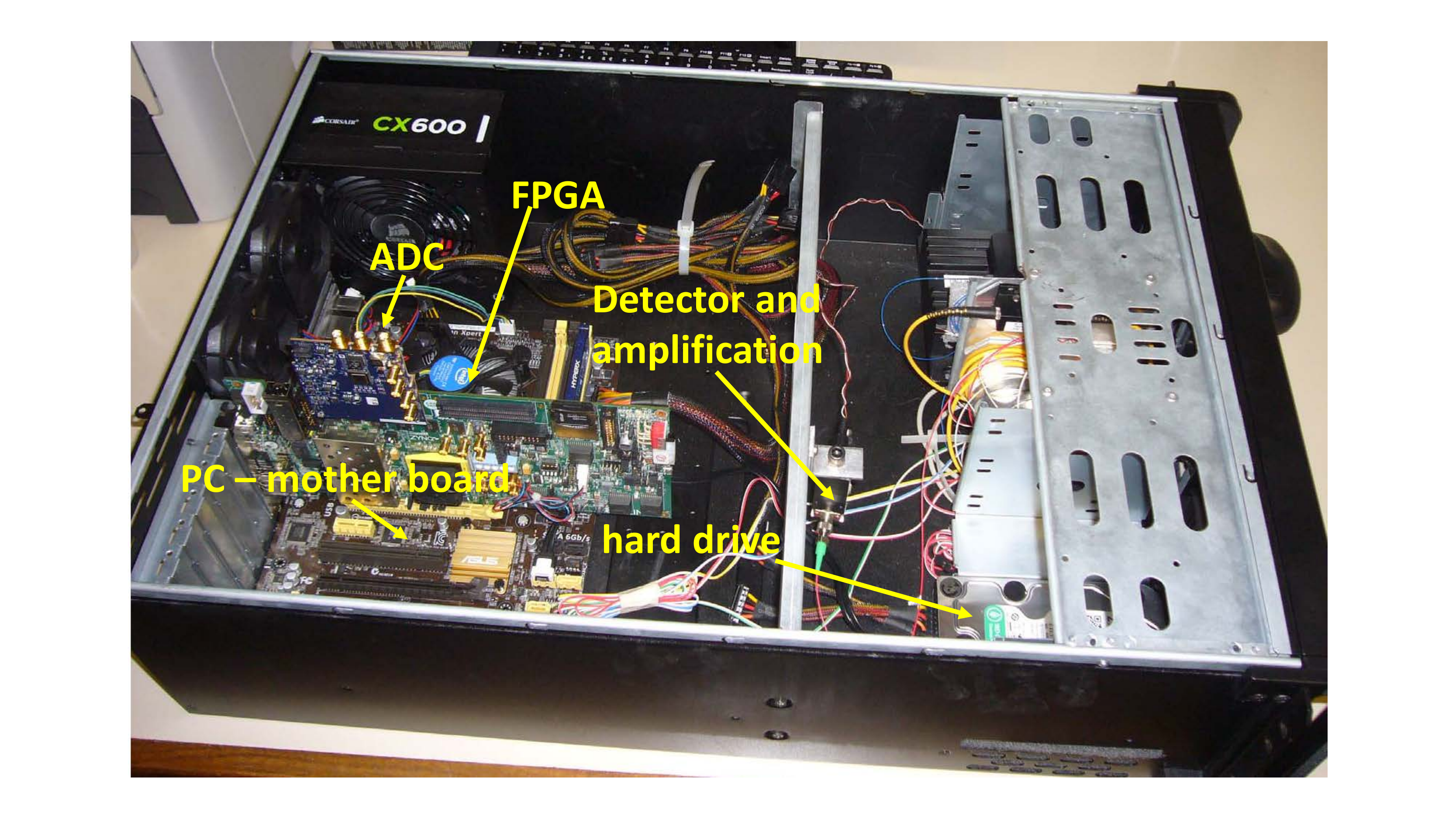}}}
\caption{ \label{PhRBGview}
A top view with some components of the platform.  The laser, detector, amplifier and hard drive are at the right side of the rack. The ADC that format analog signals from the optical amplification is connected to the FPGA for processing. A PC-motherboard provides management of several functions including a friendly graphical interface for the user. }
\end{figure}

\begin{figure}[h!]
\centerline{\scalebox{0.2}{\includegraphics{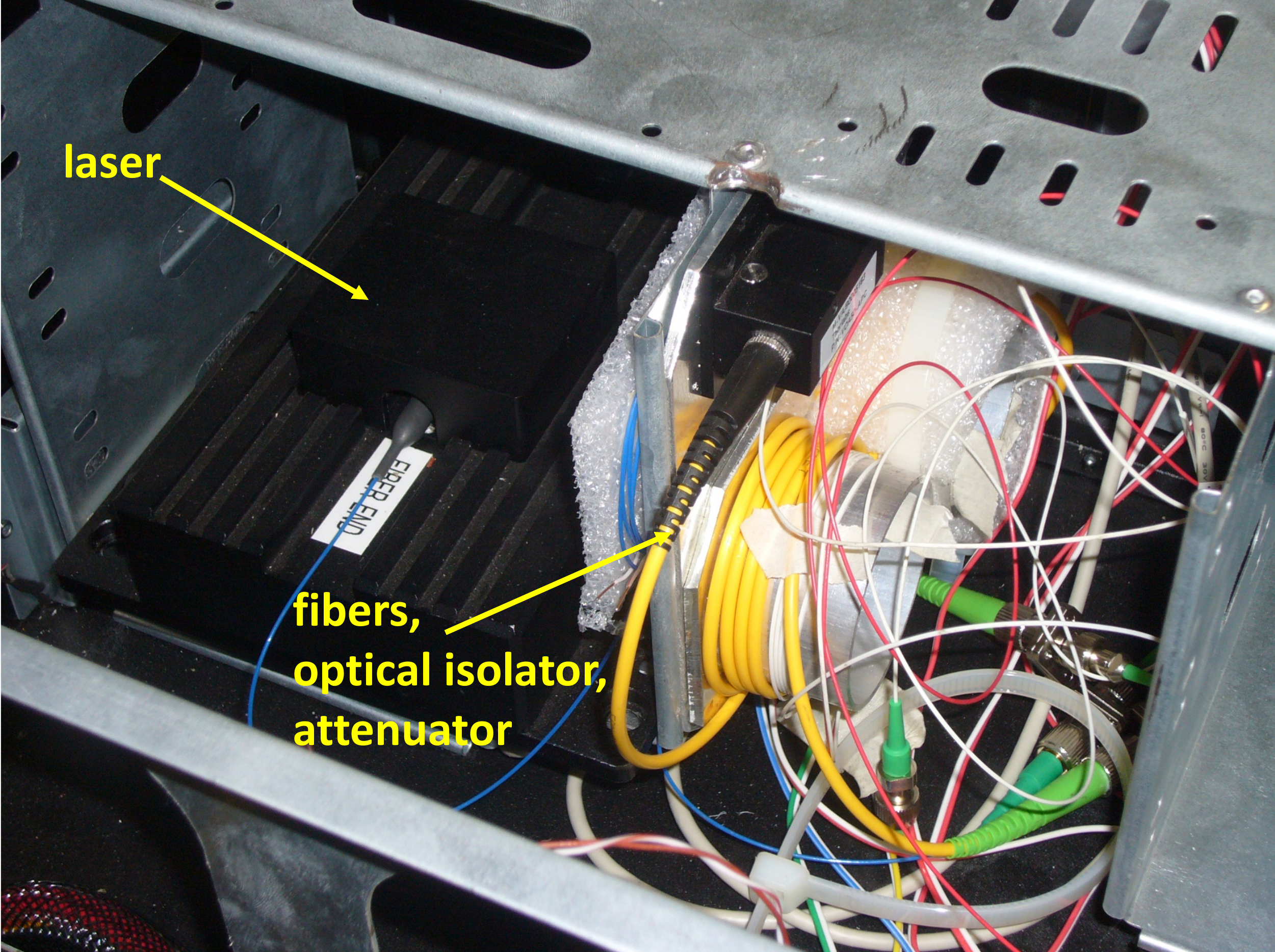}}}
\caption{ \label{PhRBGlaserFibers}
Detail of the laser location, optical isolator and attenuator. }
\end{figure}

 The PhRBG, within the platform, is seen as a rack implementation in  Fig.~\ref{PhRBGview} and some details in Fig.~\ref{PhRBGlaserFibers}.  A detailed description of the PhRBG will be published elsewhere  \cite{QuantaSEC-UFMG}. Just a brief description is presented here.

 The PhRBG is an opto-electronic device designed to generate bits continuously to supply any demand for bits at high speeds. The physical principle involved, quantum vacuum fluctuations that produce the optical shot-noise, is not bandwidth limited and the device speed can be adapted to all electronic improvements. Among the differences with other quantum random bit generators the presented device has no need for interferometry and a single detector is used. This gives a time stable operation for the system.

 The PhRBG was currently implemented with off-the-shelf components including low cost amplifiers (See G in Fig.~\ref{wireless-PhRBGblocks}). These amplifiers have a frequency dependent gain profile (a monotonous high gain at low frequencies) that introduces a low frequency bias in the bit generation. To compensate for this bias without increasing costs a Linear Feedback Shift Register (LFSR) is used in series with the bit output to produce an extra randomization. This breaks --the expectedly more rare-- long sequences of repeated bits. This process does not reduce the speed of the PhRBG.

\begin{figure}
\centerline{\scalebox{0.23}{\includegraphics{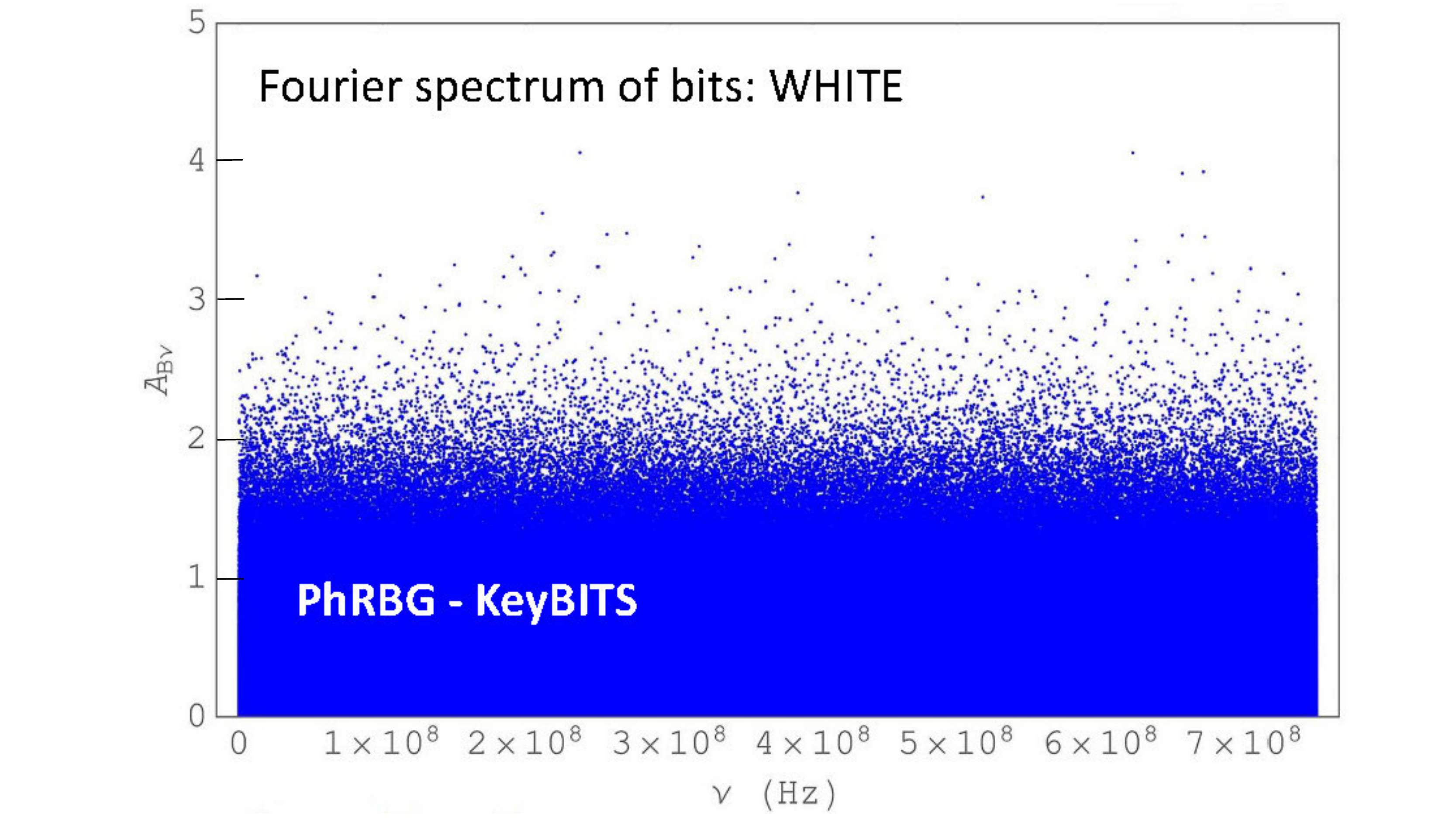}}}
\caption{ \label{PhRBGWhiteSpectrum}
 Plot of relative Fourier amplitudes $A_{\nu}$  as a function of the frequency $\nu$. Transforming $(0,1)$ sequences onto $(-1,1)$ sequences allows easy Fourier spectrum analysis that show the ``white-noise" character of the output signals. }
\end{figure}

\begin{figure}[h!]
\centerline{\scalebox{0.4}{\includegraphics{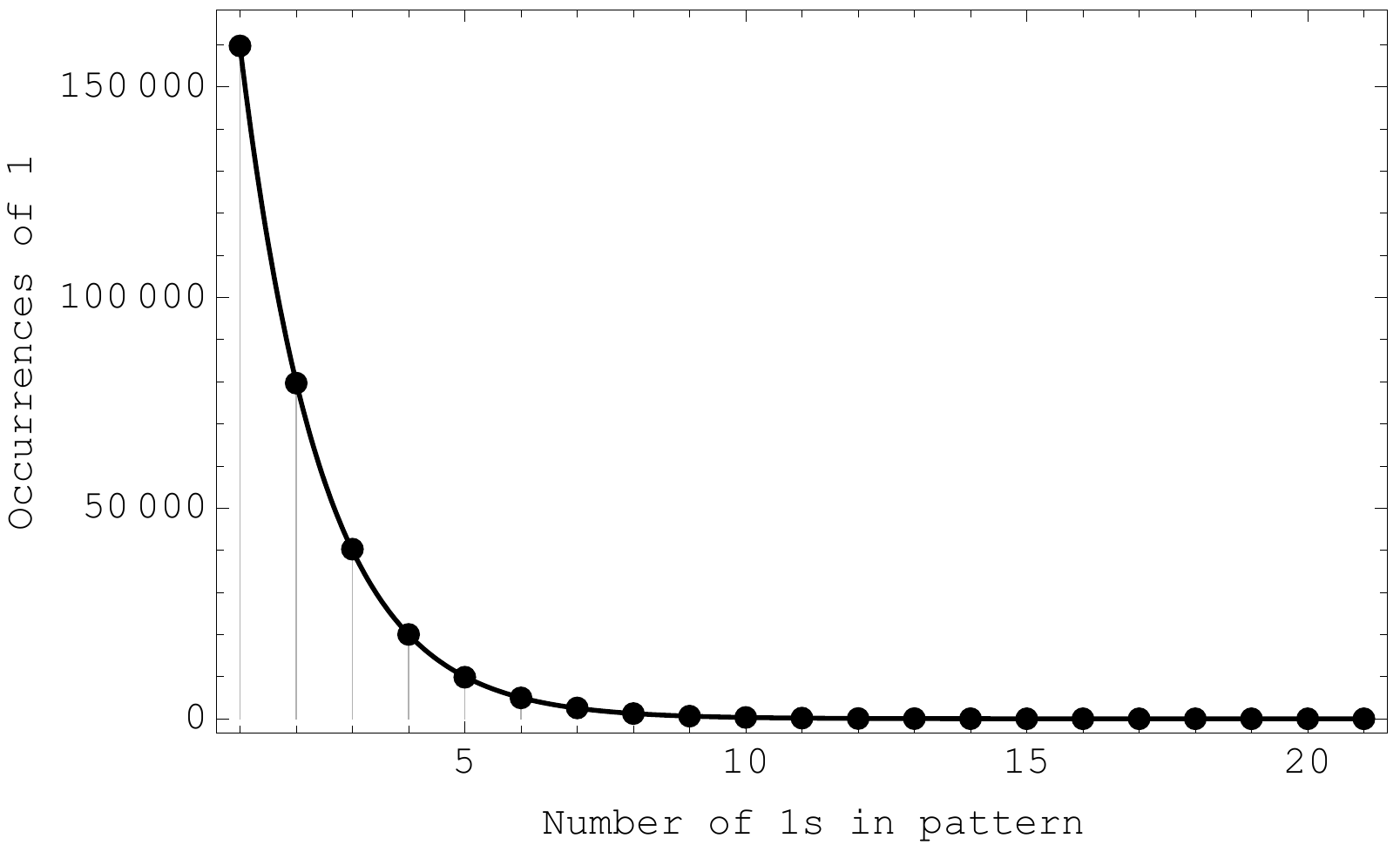}}}
\caption{ \label{pall1MillionE}
Histogram of 1s. Dots are obtained from 1,277,874 bits obtained and the solid line is the fit to $c=319018\pm 356$ and $\epsilon=-0.003\pm0.003$.}
\end{figure}

\begin{figure}[h!]
\centerline{\scalebox{0.55}{\includegraphics{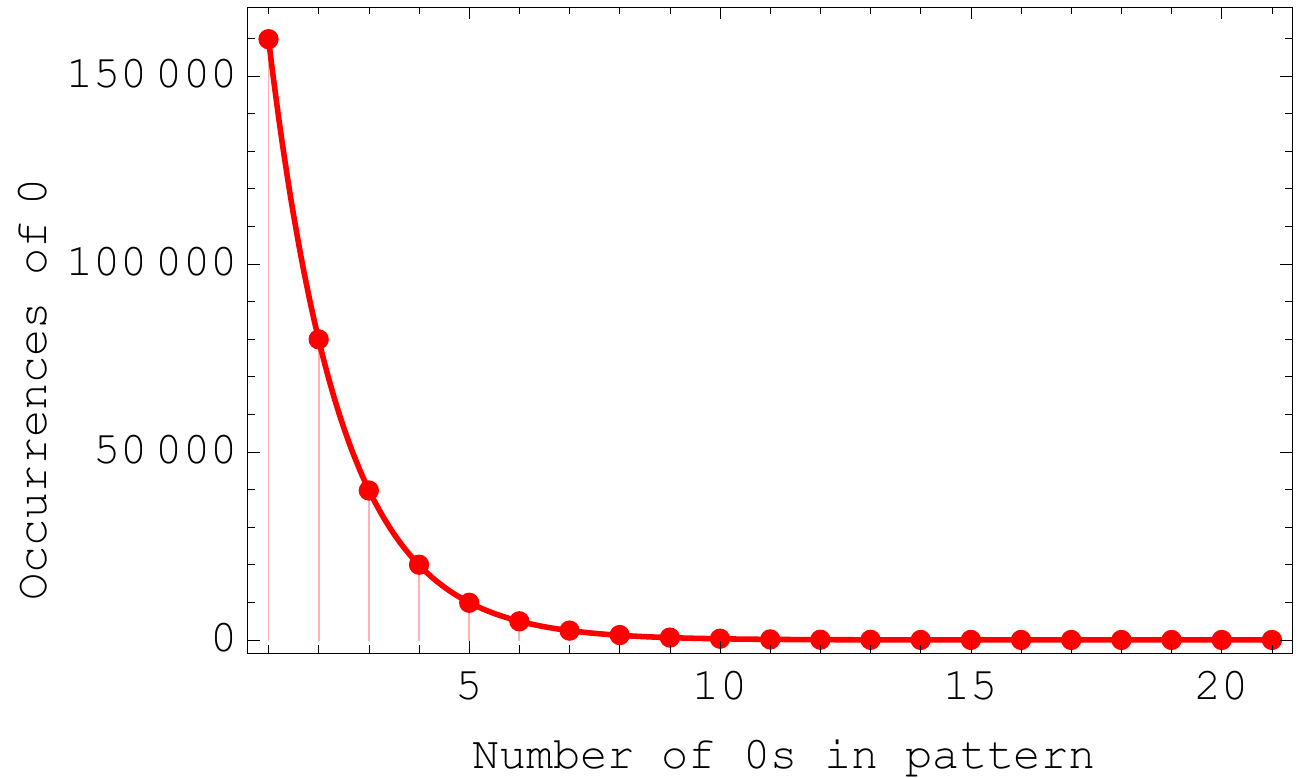}}}
\caption{ \label{pall1Million0E}
 Histogram of 0s. Dots are obtained from 1,277,874 bits obtained and the solid line is the fit to $c=319880\pm 193$ and $\epsilon=-0.003\pm0.002$. }
\end{figure}

The currently implemented PhRBG works at $\sim$2.0 Gbit/sec and passes all randomness tests to which it was submitted, including the NIST suite described in ``NIST's Special Publication 800 - A Statistical Test Suite for Random and Pseudorandom
Number Generators for Cryptographic Applications".
Besides passing conventional randomness tests, some visual information conveys the same idea. Fig.~\ref{PhRBGWhiteSpectrum} shows amplitudes of a Fourier analysis of a bit stream revealing the white spectrum character of the generated bits.
 Figs.~\ref{pall1MillionE}  and \ref{pall1Million0E} show data and the expected occurrence of random bits for a distribution where the probability to occur 0 or 1s are equal, $p=1/2$. It is expected that the probability to occur a sequence of $k$ identical bits (either 0 or 1) is $p(k)=1/2^k$.  If one changes basis 2 to basis ``e" one writes
\bea
p(k)=\frac{1}{2^k}=e^{-k \ln 2} \simeq e^{-0.693147 \: k} \:.
\eea
Data in Figs.~\ref{pall1MillionE}  and \ref{pall1Million0E} were fitted to $p(n)=c \: e^{- a n}=c\: e^{\ln 2^{1-\epsilon}n}$, where $\epsilon$ will indicate a depart from the distribution $p(k)=1/2^k$.

The raw data \cite{KeyBITS Report 3} for the histograms are given by lists $L_1$ and $L_0$:
{\small
\bea
L_1&=&\{\{1, 159676\}, \{2, 79651\}, \{3, 40253\}, \{4, 20017\}, \{5, 9864\},\nonumber \\
 && \{6,4960\}, \{7, 2567\}, \{8, 1239\},\{9, 623\}, \{10, 313\},
 \{11, 156\},\nonumber \\
 &&  \{12,59\}, \{13, 37\}, \{14, 21\}, \{15, 9\},
 \{16, 8\},\{17, 3\}, \{18, 4\},  \nonumber \\ &&\{19,
  1\}, \{20, 0\}, \{21, 0\}\}\\
L_0&=&\{\{1, 159805\}, \{2, 79964\}, \{3, 39766\}, \{4, 20021\}, \{5, 9892\}, \nonumber \\ &&
\{6, 4962\}, \{7, 2488\}, \{8, 1306\},\{9, 630\}, \{10, 336\},
 \{11, 148\},\nonumber \\ && \{12,71\}, \{13, 42\}, \{14, 10\}, \{15, 11\},
 \{16, 6\},\{17, 2\}, \{18, 0\}, \nonumber \\ && \{19,
  1\}, \{20, 1\}, \{21, 1\}\}\:.
\eea
}
One should observe that the deviation parameter $\epsilon$ is exponentially small, giving an estimate of the randomness associated with the generated bits.


\end{document}